# TRANSLATION-ROTATION COUPLING IN TRANSIENT GRATING EXPERIMENTS : THEORETICAL AND EXPERIMENTAL EVIDENCES

A.Taschin [1,2], R.Torre [1,2], M.Ricci [1,3], M.Sampoli [2,4]

1  *LENS and Dip. di Fisica, Univ. di Firenze, largo E.Fermi 2, 50125 Firenze, Italy;*
2  *INFM, Unità di Firenze, largo E.Fermi 2, 50125 Firenze, Italy;*
3  *Dip. di Chimica Fisica, Univ. della Basilicata and INFM, Unità di Napoli, Italy;*
4  *Dip. di Energetica, Univ. di Firenze, via S.Marta, Firenze, Italy.*

and

C.Dreyfus[1], R.M.Pick[2]

1  *PMC, UMR 7602, B.P. 77, Univ. P. et M. Curie, 4 Place Jussieu, F-75005, Paris, France,*
2  *L.M.D.H., UMR 7603, B.P.86, Univ. P. et M. Curie, 4 Place Jussieu, F-75005, Paris, France.*

## ABSTRACT

The results of a Transient Grating experiment in a supercooled molecular liquid of anisotropic molecules and its theoretical interpretation are presented. These results show the existence of two distinct dynamical contributions in the response function of this experiment, density and orientation dynamics. These dynamics can be experimentally disentangled by varying the polarisation of the probe and diffracted beams and they have been identified and measured in a Heterodyne Detected experiment performed on *m*-toluidine. The results of the theory show a good qualitative agreement with the measurements at all temperatures.

# 1 Introduction

When an intense ultra-short laser pulse of near-infrared wave-length passes through a molecular liquid of anisotropic molecules, it produces three main effects: (i) it is partly absorbed, generating a local heating, (ii) it creates an instantaneous electrostrictive pressure and (iii) it partly orients the molecules along its electric field through coupling with the anisotropic part of the molecular polarisability tensor (Optical Kerr Effect-OKE). The local pressure induced by the two first effects generates a density perturbation which propagates through the liquid. The resulting non-uniform velocity field generates a strain rate that couples to the molecular orientations. Molecules are thus oriented both directly by the electric field and indirectly by the density perturbation, and both effects generate a local optical anisotropy.

In a recent paper, Hinze et al. [1] studied the orientation dynamics of such a supercooled molecular liquid by means of a time-resolved experiment, called DIHARD (Density Induced, Heterodyne Amplified, Rotational Dynamics), based on the laser-matter interactions summarised above. They induced an optical anisotropy in the sample by a short laser pulse and probed it by the change of polarisation, measured by a heterodyne detection (HD), of a second laser beam when it passes through the liquid in a region not illuminated by the pump. Hinze et al. showed that they could differentiate between the two origins of the anisotropy by selecting proper polarisations of the pump and the probe beams in a heterodyne detected (HD) experiment. They proposed that the HD-OKE signal, (produced by the electric field orientation) would be proportional to the time derivative of the DIHARD signal (produced by the strain rate). They finally suggested that similar and complementary results could be obtained using a Transient Grating (TG) technique of the type described by Nelson and co-workers [2]. Yet, the interpretation of such a new experiment required also new theoretical developments because the theory of T.G. experiments presented in [2] applied only to fluids formed of shapeless particles

This letter reports the first HD-TG experiment with polarisation selection performed on a molecular supercooled liquids, *m*-toluidine, as well as the results of a complete phenomenological theory that takes into account the anisotropic character of the molecules of the supercooled liquid. The strain rate orientation effect, corresponding to the DIHARD signal, is experimentally identified and is shown to persist down to the glass transition temperature in agreement with the theory. The HD-TG results clearly show that the OKE signal has an origin largely different from the DIHARD signal, which partly explains why they have very different intensities.

## 2 Transient Grating experiments and their theoretical description

In a TG experiment, two coherent light pulses (the pump), with a wavelength $\lambda$ and wave-vectors $\vec{q}_1$ and $\vec{q}_2$ interfere inside the liquid, $\vec{q}_2$ forming a small angle $\theta$ (~6 degrees for q=0.63µm$^{-1}$) with $\vec{q}_1$. These two beams generate an interference field, $\vec{E}(\vec{r},t)$, which is a standing wave with a wave vector $\vec{q}$, directed along the external bisector of $\vec{q}_1$ and $\vec{q}_2$ with $|\vec{q}| = \frac{4\pi}{\lambda}\sin\theta/2$. This electric field generates a grating of wave-vector $\vec{q}$ in the material through the three distinct mechanisms described in the introduction, and this grating is probed by a third light beam that is diffracted by the grating. In the limit of an impulsive excitation, the HD signal, $S^{HD}(t)$, can be expressed as [3]:

$$S^{HD}(t) \propto \delta\varepsilon_{ij}(q,t) \propto R_{ijkl}(q,t) F_{kl} \qquad (1)$$

where $F_{kl}$ represents the exciting force(s) produced by the two beams of the pump, $k$ and $l$ being the Cartesian coordinates of the electric fields of these beams, while $i$ and $j$ refer to the electric fields of the probe and the diffracted beams respectively [3]. $R_{ijkl}(q,t)$ is the response function of the system that defines the dynamical properties of the experimental observable, the local dielectric tensor $\delta\varepsilon_{ij}(q,t)$. In the experiments performed here, the directions of the four electric fields were such that $i=j$ and $k=l$ : only $R_{\alpha\alpha\beta\beta}$ component were detected. The measured signal can thus be written as $S^{HD}_{aabb} \propto R_{aabb}$, $\alpha$ and $\beta$ being here directions either perpendicular to the scattering plane, $(\alpha,\beta \equiv v)$, or in this plane, $(\alpha,\beta \equiv h)$.

The four corresponding $R_{aabb}$ functions can be obtained by generalising a phenomenological set of equations recently proposed in [4] to describe the equilibrium dynamics of supercooled molecular liquids with axially symmetric molecules. In the present case, a first equation describes the linear coupling between the dynamics of a traceless tensor, $Q_{ij}$, associated with the local molecular orientation probability distribution function and the local, traceless, shear strain rate tensor, $\tau_{ij}$. A second equation expresses the dynamics of the mass density, $\rho$, when subjected to retardation effects related to both variables, $\rho$ and $Q_{ij}$, as well as to a pressure gradient. The last equation describes the thermal evolution of the system. This equation was not taken into account in [4] and will be reported in a forthcoming paper [5] together with the detailed treatment of this theoretical model. Within this model, the formation and time evolution of the TG is obtained by adding a source term to each of these equations, in a manner similar to the previous treatment of Yang and Nelson [2] in which the existence of the orientational variable was not included. The sources, or forcing terms, are defined by the coupling of the interference field, $\vec{E}(\vec{r},t)$, with the relevant variables of the system, these sources being the three effects described in the Introduction, each of them having a strength proportional to the excitation forcing term, $F(\vec{r},t) \propto |\vec{E}_0| \cos(\vec{q}.\vec{r}) d(t)$. Performing the Laplace Transforms (L.T.) of these equations, the local temperature of the liquid is first obtained. Then, $\rho(q,\omega)$ and $Q_{ij}(q,\omega)$ are derived ; they depend linearly on each of these three sources, and in particular on the polarisation of $\vec{E}(\vec{r},t)$. Furthermore, the experimental observable, the fluctuations of the local dielectric tensor, is linearly connected to the two dynamical variables, $r$ and $\overline{\overline{Q}}$, by:

$$\delta\varepsilon_{ij}(\vec{r},t) = a\delta_{ij}\delta\rho(\vec{r},t) + bQ_{ij}(\vec{r},t) \qquad (2)$$

where $a$ and $b$ are constants. Selecting the appropriate polarisation of the excitation and probe beams, one finally gets :

$$R_{aabb}(q,t) \approx \frac{1}{2\pi} \int_0^\infty \cos wt \quad Re(R_{aabb}(q,w)) \; dw \qquad (3)$$

where the function $R_{aabb}(q,w)$ is the corresponding response function in the frequency space. The theory sketched above leads to :

$$R_{aabb}(q,w) = FD(w)b(1+3e_{ex}e_p) + \left\{ \frac{H}{1+iwt_h} - q^2[K + F(3e_{ex}-1)r(w)] \right\} P(w)[a + b'(3e_p - 1)r(w)]$$

(4)

where α stands for v ($\varepsilon_p = 1$) or h ($\varepsilon_p = -1$) and β stands for v ($\varepsilon_{ex} = 1$) or h ($\varepsilon_{ex} = -1$). Let us describe here the physical meaning of the different contributions present in this equation.

The first term on the r.h.s. of Eq. 4 is the Optical Kerr Effect contribution (HD-OKE) and it contains three factors : (i) coefficient $F$, which represents the coupling of the electric field of the pump to the molecular polarisability; (ii) the propagator $D(w)$ of the molecular librations, with $D(\omega) = [\omega_1^2 + \omega\Gamma(\omega) - \omega^2]^{-1}$ (where $\omega_1$ is the frequency of the local librations of the molecules and $\Gamma(\omega)$ is the L.T. of the orientational relaxation function associated with them); (iii) coefficient $b$, which makes it clear that this signal is detected through the corresponding change of $Q_{ij}$.

The second term of the r.h.s. of Eq. 4 is also the product of three factors. The first factor, between braces, is the sum of the three terms originating from the three sources: the first term arises from the energy, $H$, absorbed by the liquid which diffuses in the liquid with a relaxation time $\tau_h = C_v / \lambda q^2$, where $C_v$ is the specific heat and $\lambda$ the thermal conductivity; the second term represents the contribution of the electrostriction effect, $K$, while the last term, again proportional to $F$, is produced by the coupling of the pump electric field, $\vec{E}(\vec{r},t)$, with the anisotropic part of molecular polarisability; it generates longitudinal phonons through the rotational translation coupling function, $r(\omega)=\omega m(\omega)D(\omega)$, where $m(\omega)$ is the L.T. of the relaxation function associated with the coupling of the shear strain rate to the dynamics of $Q_{ij}$. The second factor is the usual longitudinal phonon propagator: $P(\omega) = [q^2 c_0^2 + \omega \eta_L(\omega)\rho_m^{-1} - \omega^2]^{-1}$, where $c_0$ is the relaxed longitudinal velocity, $\eta_L(\omega)$ is the frequency dependent longitudinal viscosity (see Eq. A14 of [4]) and $\rho_m$ is the average mass density. The last factor connects the experimental observable to both the density ($a$ term) and to the molecular orientation variable ($b'\, r(\omega)$ term, where the coefficient $b'$ is proportional to $b$).

## 3 Experimental results

The present experiment has been performed with a laser system and an optical set-up which will be described elsewhere [6]. This experiment consisted in an extensive study of an organic glass-forming liquids, $m$-toluidine, which had been purified by repeated distillation under vacuum, starting from products obtained from Merck and, later, kept in a quartz cell. The measurements were performed from 298 K, well above the melting temperature, $T_m$= 243 K [7], down to 180 K, i.e. below the glass transition temperature, $T_g \sim 187$ K. A 0.1 K stability was achieved using a Cryogenics cryostat in conjunction with a Lake Shore control system.

Measurements were performed for the four polarisation configurations: $S_{vvvv}$, $S_{hhvv}$, $S_{vvhh}$, and $S_{hhhh}$, [8]. Whatever the temperature, the HD-TG signal does not show any significant difference in profile or in intensity by changing the direction of polarisation of the pump : $S_{vvhh} \approx S_{vvvv}$ and $S_{hhhh} \approx S_{hhvv}$, (see Fig.1). On the contrary, the TG signal profile depends on the direction of polarisation of the probe, $S_{hhvv} \neq S_{vvvv}$, and the difference extends to higher time domains with decreasing temperature. These differences start to appear below 230 K, the temperature at which the signature of the structural relaxation begins to mix the heat diffusion signal : the $S_{hhvv}$ signal becomes less intense at short and intermediate times than the $S_{vvvv}$ signal. At still lower temperatures, the difference extends to the whole time domain while the two signal become identical in shape, Fig.2. We checked these polarisation effects on other glass-formers : very similar results were obtained for salol while, vice versa, substantially no polarisation effects were detected in *o*-ter-phenyl (OTP) and glycerol.

## 4 Interpretation and Analysis

Let us analyse the preceding experimental results within the framework of Eq. 4. These results imply that the *m*-toluidine response function is insensitive to the value of $\varepsilon_{ex}$, which defines the polarisation of the pump, but sensitive to the value of $\varepsilon_p$, i.e. to the probe and diffracted beam polarisations. As $\varepsilon_{ex}$ is only related to the coefficient F of Eq. 4, that coefficient can be neglected for the interpretation of the experiments performed here [9]. Due to this neglect, the role of the two dynamical variables on the detection mechanism can be disentangled in Eq. 4 by performing the two linear combinations :

$$\frac{1}{3}(2S_{vvvv} + S_{vvhh}) \approx \frac{a}{2\pi} \int_0^\infty \cos\omega t \ P(\omega) \left( \frac{H}{1+i\omega\tau_h} - q^2 K \right) d\omega \qquad (5a)$$

$$\tfrac{1}{2}(S_{vvvv} - S_{vvhh}) \approx \frac{b'}{2\pi} \int_0^\infty \cos\omega t \ r(\omega) \ P(\omega) \left( \frac{H}{1+i\omega\tau_h} - q^2 K \right) d\omega \qquad (5b)$$

The l.h.s. of Eq. 5a is the isotropic part of the TG signal. This response is analytically described by its r.h.s. and is the density response function, according to the general hydrodynamics equations [2]. Similarly, the l.h.s. of Eq. 5b is the anisotropic part of the response function, detected through the orientation of the molecules. It is the TG equivalent of the DIHARD signal of [1], corresponding to similar linear combinations of the experimental signals. The comparison between the r.h.s. of Eqs. 5a and 5b shows that the difference between the two terms arises, apart from their overall intensity, only from the extra $r(\omega)$ factor in Eq. 5b.

In Fig. 3, we report the isotropic and anisotropic HD-TG signals obtained in our experiment for $q=0.63$ µm$^{-1}$ at the three temperatures T=220 K, T=208 K and T=195 K. The signals have clearly different shapes for the two highest temperatures but only different intensities at 195 K. In a first step towards the validation of our theoretical description of these signals, we now compare the results obtained in the present study and, in particular, those shown in Fig. 3 with their predicted form (Eqs. 5). For this comparison, we need to make precise the form of the relaxation functions entering into the r.h.s. of these equations and to give the numerical values chosen for the different corresponding parameters. For $P(\omega)$, we expressed $wh_L(w)$ as:

$$wh_L(w) = d_0^2\left(1 - \frac{1}{(1+iwt_L)^b}\right) + iwg_0$$

and, in agreement with previous light scattering analyses of *m*-toluidine [4, 10], we took $g_0= 0.09 \cdot 10^9$s$^{-1}$, $b= 0.6$, $c_0$ and $d_0$ being derived from the experimental values obtained in [10]. For each temperature, the longitudinal relaxation time, $t_L$, and the heat diffusion relaxation time, $t_h$, were adjusted to obtain a qualitative agreement with the spectra of Fig. 3. The instrumental response function of the set-up was measured independently and convoluted with the theoretical expressions (Eqs. 5) to produce the results shown in Fig. 4.

The isotropic (density) signal (Eq. 5a) is the sum of two terms which are, following the terminology of [2], the ISTS (Impulsive Stimulated Thermal Scattering) term, proportional to $H$, and the ISBS (Impulsive Stimulated Brillouin Scattering) term, proportional to $K$. Their shape and thermal evolutions, described, e.g., in [11], are very characteristic and different one from the other: the ISTS signal corresponds, at short times (here, for t < or ~ $10^2$ ns), to the decay of the phonon and, at longer times, to the thermal diffusion process, possibly mixed (as it is the case for the three temperatures considered here) with the longitudinal viscosity relaxation process. Conversely, the ISBS signal is only related to the phonon relaxation process, and never extends to longer times. Furthermore, for small values of $q$, due to the $q^2$ factor in front of $K$, the ISTS term generally dominates the isotropic signal [6, 11-13] and this turns out to be also the case of *m*-toluidine. The HD-TG signals shown in Fig.2 correspond to temperatures at which the relaxation time of the phonons, ($\frac{1}{g_0}$), is shorter than $t_L$. In that case, $t_L$ and $t_h$, are the two parameters which mostly determine the shape of the ISTS term at times much longer than $\frac{1}{g_0}$; this allows to obtain rather precise values for $t_L$ (cf. [6, 11-13]). The isotropic theoretical signals shown in Fig.4 are the sums of the two preceding terms with a small, adjusted, ISBS contribution ; the values of $t_L$ we had to use for those three curves favourably compare with the corresponding times obtained in [10].

To compute anisotropic response functions, Eq. 5b, we neglected the frequency dependence of $D(w)$ (see [4]), assumed for $wm(w)$ a Cole-Davidson form with the same value of $b$ used in $wh_L(w)$ and postulated for the corresponding relaxation times, $t_m$, the temperature independent relationship $t_m = at_L$. We chose a value of $a$ equal to 3, a plausible estimate, which relies on the hypothesis that $t_m$ would have a value close to that of the rotational relaxation time [10]. We computed the r.h.s. of Eq. 5b for values of $t_L$ ranging from $10^{-2}$ ns to $10^6$ ns. For all temperatures, the term originating from the electrostrictive effect is approximately proportional to its ISBS counterpart in Eq. 5a, the proportionality factor being less than unity ; its role is thus always very small. The term proportional to $H$ is also rather weak at high temperature (shortest values of $t_L$) with no long time tail : this explains that the anisotropic signal is not detected above 230 K. Conversely, for lower temperatures, the intensity of the term proportional to $H$ becomes larger, the signal extends up to times longer than $t_L$ and its shape differs from the ISTS signal for the two highest temperatures of Fig. 3, for which $\frac{1}{g_0} \ll t_L \ll t_h$. Finally, at 195 K, where $t_h \ll t_L$, the isotropic and anisotropic signals have an identical shape but different intensities, $r(w)$ becoming essentially frequency independent for such a large $t_m$ value. The anisotropic (rotational) contributions computed with the help of the numerical values obtained in the preceding paragraph are also shown in Fig. 4. They exhibit a remarkable qualitative agreement with the experimental results of Fig.3.

**Conclusion**

Thanks to polarisation effects, the present HD-TG experiment is able to measure selectively different aspects of the relaxation dynamics of a supercooled molecular liquid of anisotropic molecules. More precisely, it allows to isolate the isotropic (density) and the anisotropic (orientational) contributions to the signal with an extremely high signal/noise ratio. Furthermore the analysis performed above shows that the origin of the orientational, or DIHARD, signal is to a very large extent the first term of the r.h.s. of Eq. 5b, in which there is a complex interplay between the roles of $t_L$, $t_m$ and $t_h$. There is thus very little relationship between the DIHARD and the HD-OKE signals. Conversely, the simultaneous analysis of the isotropic and anisotropic contributions should lead to reliable values of $t_L$ and $t_m$ in the

$\frac{1}{g_0} \ll t_L \ll t_h$ regime. That would correspond to the first measurement of $t_m$ or, more precisely, to the simultaneous measurement of $t_L$ and $t_m$ allowing for a precise determination of this ratio at various temperatures. Quantitative fits of TG data on the basis of the analysis presented here are in progress.

This work was supported by INFM through the project TREB-Sez.C-PAISS1999 and by the Commission of the European Communities through the contract N° HPRI--CT1999-00111 and by MURST.

-

**Figure Captions**

Figure 1. Signal measured at $q=1$ μm$^{-1}$ and T=220 K, for the four different polarisation configurations. $S_{vvvv}$ and $S_{vvhh}$ signals are shown in the upper graph, $S_{hhhh}$ and $S_{hhvv}$ signals are shown in the lower graph : the two left indices refer to the polarisation of the probe and diffracted beams and the two right indices to the polarisation of the two beams forming the pump. Signals with identical probe and detection polarisations but different pump polarisations do not show any significant difference.

Figure 2. The $S_{vvvv}$ and the $S_{hhvv}$ signals for $q=0.63$ μm$^{-1}$ at 220 K, 208 K and 195 K.

Figure 3. The isotropic (density, $1/3[2S_{vvvv}+S_{hhvv}]$) and anisotropic (rotational, $½[S_{vvvv}-S_{hhvv}]$) signals computed from the results shown in Figure 2.

Figure 4. The isotropic (density) and anisotropic (rotational) response functions computed from their theoretical expressions (Eqs. 5a and b), using numerical values adapted for the three temperatures of Figure 2, the other numerical values being deduced from [10] or defined in the text.

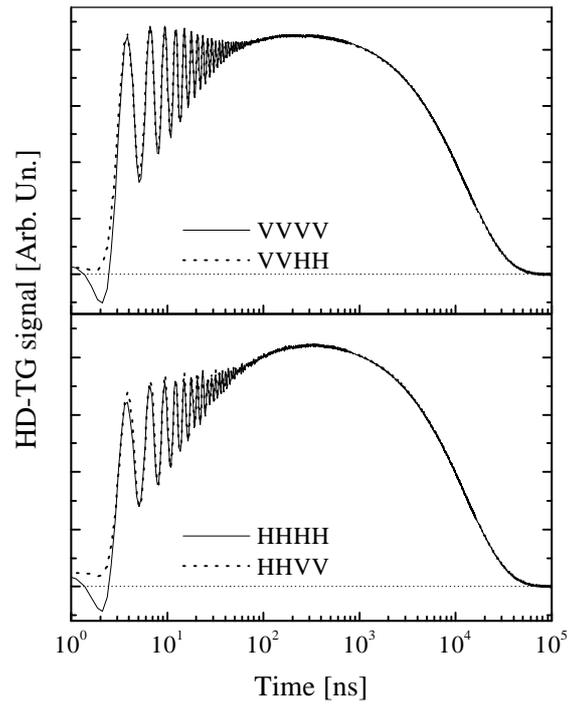

Figure 1    A. Taschin et al.

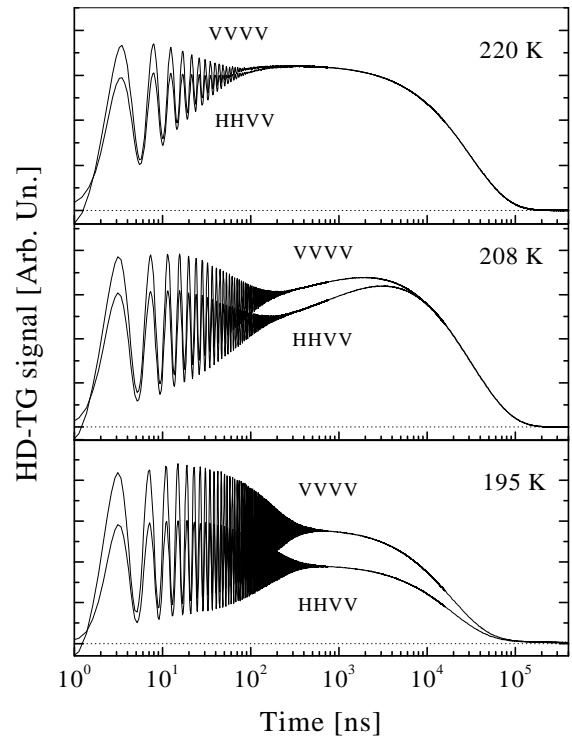

Figure 2

A. Taschin et al.

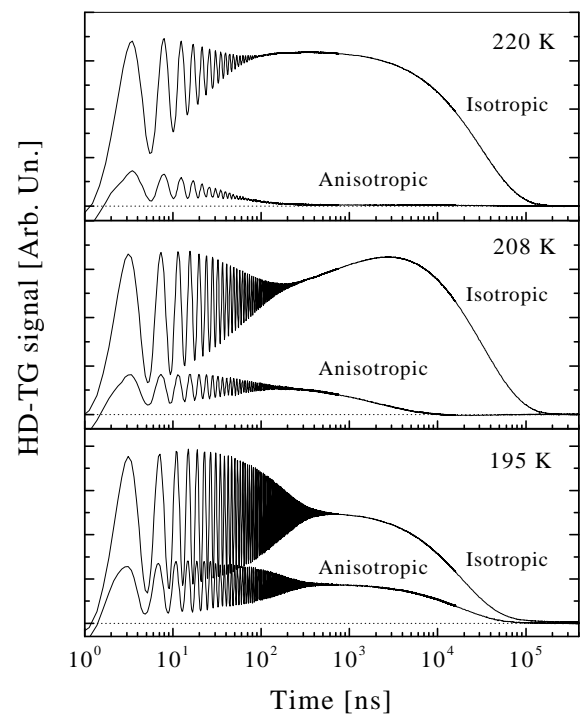

Figure 3　　　　　　　　　　　　　　　　　A. Taschin et al.

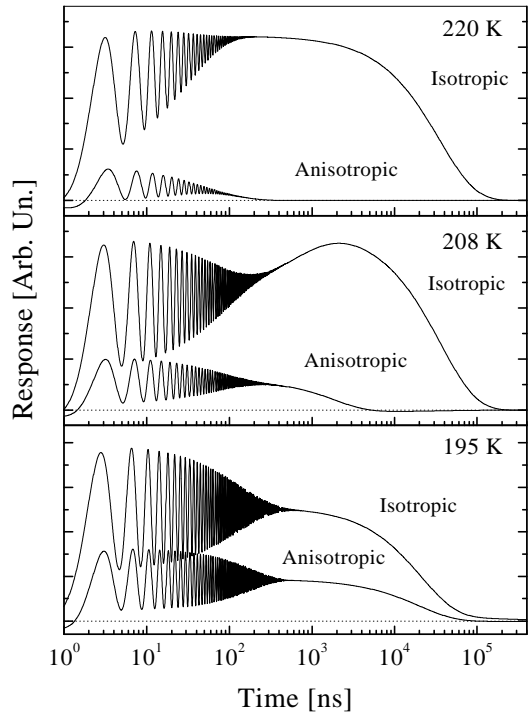

Figure 4 A. Taschin et al.